\documentclass[aps,reprint,groupedaddress,showpacs]{revtex4-1}

\usepackage[utf8]{inputenc}
\usepackage{graphicx}
\usepackage{amsmath}
\usepackage{amssymb}
\usepackage{enumerate}

\def \M {\mathcal{M}}
\def \R {\mathcal{R}}

\begin{document}

\title{Effective spacetime geometry of graviton condensates in $f(\R)$ gravity}

\author{Andy Octavian Latief}
\email{latief@fi.itb.ac.id}
\affiliation{Physics of Magnetism and Photonics Research Division, Faculty of Mathematics and Natural Sciences, Institut Teknologi Bandung, Jl.~Ganesha no.~10 Bandung, Indonesia, 40132}

\author{Fiki Taufik Akbar}
\email{ftakbar@fi.itb.ac.id}
\affiliation{Theoretical Physics Laboratory, Theoretical High Energy Physics Research Division, Faculty of Mathematics and Natural Sciences, Institut Teknologi Bandung, Jl.~Ganesha no.~10 Bandung, Indonesia, 40132}

\author{Bobby Eka Gunara}
\email{bobby@fi.itb.ac.id (Corresponding author)}
\affiliation{Theoretical Physics Laboratory, Theoretical High Energy Physics Research Division, Faculty of Mathematics and Natural Sciences, Institut Teknologi Bandung, Jl.~Ganesha no.~10 Bandung, Indonesia, 40132}

\begin{abstract}
We consider a model of Bose-Einstein condensate of weakly interacting off-shell gravitons in the regime that is far from the quantum critical point. Working in static spherically symmetric setup, recent study has demonstrated that the effective spacetime geometry of this condensate is a gravastar. In this paper we make three generalizations: introducing a composite of two sets of off-shell gravitons with different wavelength to enable richer geometries for the interior and exterior spacetimes, working in $f(\R)$ gravity, and extending the calculations to higher dimensions. We find that the effective spacetime geometry is again a gravastar, but now with a metric which strongly depends on the modified gravity function $f(\R)$. This implies that the interior of the gravastar can be de Sitter or anti--de Sitter and the exterior can be Schwarzschild, Schwarzschild--de Sitter, or Schwarzschild--anti--de Sitter, with a condition that the cosmological constant for the exterior must be smaller than the one for the interior. These geometries are determined by the function $f(\R)$, in contrast to previous works where they were selected by hand. We also presented a new possible value for the size of the gravastar provided a certain inequality is satisfied. This restriction can be seen manifested in the behavior of the interior graviton wavelength as a function of spacetime dimension.
\end{abstract}

\maketitle

\section{Introduction}

According to classical general relativity, black hole is a very dense object with a curvature singularity at the origin and a coordinate singularity known as the event horizon at some radius. Any particle, and even light, cannot escape from the black hole once it enters the event horizon, making the black hole interior inaccessible to outside observers. Although the existence of curvature singularity is already a controversial issue, the picture of black hole becomes more problematic when quantum effects are included (see Refs.~\cite{Wald-thermodynamics-2001,Brustein-Medved-non-2019} for reviews). For example, Hawking discovered that black holes can evaporate due to a mechanism later known as the Hawking radiation \cite{Hawking-particle-1975}. This implies that an information in the form of a pure quantum state can transform into a mixed state, which contradicts the unitarity of quantum mechanics.

Several ideas have been proposed to remove these problems or even to change drastically the physical picture inside the black hole interior \cite{Brustein-Medved-non-2019}. One of them is the gravastar, which stands for the gravitational vacuum star, proposed by Mazur and Mottola \cite{Mazur-Mottola-gravitational-2002,Mazur-Mottola-dark-2004,Mazur-Mottola-gravitational-2004}. The idea is that when an astronomical object undergoes a gravitational collapse, a phase transition occurs at the expected position of the event horizon to form a spherical thin shell of stiff fluid with equation of state $p = \rho$. The interior is a de Sitter (dS) condensate phase obeying $p = -\rho$, while the exterior is a Schwarzschild vacuum obeying $p = \rho = 0$. There are also other proposals to generalize the interior and exterior regions of a gravastar, ranging from changing only the interior to an anti--de Sitter (AdS) spacetime \cite{Visser-Wiltshire-stable-2004} and a Born-Infeld phantom \cite{Bilic-et-al-born-2006} to even changing both the interior and exterior regions by including also the case of AdS for the interior as before and generalizing the Schwarzschild exterior to Schwarzschild--de Sitter (Sch-dS), Schwarzschild--anti--de Sitter (Sch-AdS), and Reissner-Nordstr\"{o}m spacetimes \cite{Carter-stable-2005}. These varieties of gravastars have been proved to be stable under radial perturbations by the respective authors. Especially for the case where the interior is dS and the exterior is Sch-dS, there is a relation between the cosmological constants of the interior and the exterior regions: the latter must be smaller than the former \cite{Chan-et-al-how-2009}.

Another idea to solve black hole paradoxes is a proposal by Dvali and G\'{o}mez that black holes are Bose-Einstein condensates (BEC) of weakly interacting gravitons at the critical point of a quantum phase transition \cite{Dvali-Gomez-landau-2012,Dvali-Gomez-black-2012,Dvali-Gomez-black-hole-1-N-hair-2013,Dvali-Gomez-black-hole-quantum-N-portrait-2013,Dvali-Gomez-black-2014,Dvali-Gomez-quantum-2014}, which happens when $\alpha N = 1$, with $\alpha$ the dimensionless quantum self-coupling of gravitons and $N$ the number of gravitons \cite{Dvali-Gomez-black-hole-quantum-N-portrait-2013,Dvali-et-al-black-2015}. They explained that even a macroscopic black hole is a quantum object and therefore treatments using semiclassical reasoning is not adequate; one has to use full quantum treatments to avoid paradoxes. The Bekenstein entropy and Hawking radiation now have natural explanations; the former is quantum degeneracy of the condensate at the quantum critical point \cite{Dvali-Gomez-black-hole-quantum-N-portrait-2013} and the latter is quantum depletion and leakage of the condensate \cite{Dvali-Gomez-landau-2012,Dvali-Gomez-black-2014}.

In this paper, we will consider a model of Bose-Einstein condensate of weakly interacting off-shell gravitons in the regime that is far from the quantum critical point, namely where $\alpha N < 1$. (For different approach and context, see Refs.~\cite{Alfaro-et-al-bose-2017,Alfaro-et-al-condensates-2019}.) Since the language of microscopic graviton condensate should translate into the geometrical language of general relativity in the classical regime, one may ask about the effective spacetime geometry that this condensate generates. Working in the static spherically symmetric setup, Cunillera and Germani in Ref.~\cite{Cunillera-Germani-gross-2018} adapted the derivation of the Gross-Pitaevskii (GP) equation for ordinary BEC to this graviton condensate, namely by varying the condensate energy obtained from the Arnowitt-Deser-Misner (ADM) formalism while the number of gravitons is kept fixed. They found that the interior of the condensate is described by the dS spacetime while the exterior is described by the Schwarzschild spacetime, which is analogous to the picture of gravastar explained earlier. Therefore, this method provides a bridge between the theory of graviton condensates and the model of gravastars \cite{Brustein-Medved-non-2019}.

We will perform three generalizations to the work outlined in Ref.~\cite{Cunillera-Germani-gross-2018}:
\begin{enumerate}[(i)]
    \item introducing a composite of two sets of off-shell gravitons with different wavelength to enable richer geometries for the interior and exterior spacetimes,
    \item working in $f(\R)$ gravity,
    \item and extending the calculations to higher dimensions.
\end{enumerate}
In Sec.~\ref{effective-metrics} we will derive the Gross-Pitaevskii equations governing the effective metric of the graviton condensate. The combination of points (i) and (ii) above will enable us to have dS and AdS spacetimes for the interior region and Schwarzschild, Sch-dS, and Sch-AdS spacetimes for the exterior region. However, unlike previous works, these spacetimes are not selected by hand; here their cosmological constants are determined by the modified gravity function $f(\R)$. In Sec.~\ref{size-interior-condensate}, following Ref.~\cite{Cunillera-Germani-gross-2018}, we will discuss a method to determine the size of the gravastar and demonstrate that the generalization to the $f(\R)$ gravity again gives us richer results compared to the case of ordinary gravity. Then in Sec.~\ref{special-cases} we will study some special cases of interior and exterior geometries, starting from the conventional case of dS interior and Schwarzschild exterior, followed by the case of dS interior and Sch-(A)dS exterior, and completed by the case of AdS interior and exterior. The behavior of the interior graviton wavelength as a function of spacetime dimension is studied in each of these cases. The paper is then concluded in Sec.~\ref{conclusions}.

\section{The effective metric}

\label{effective-metrics}

Consider a condensate of weakly interacting off-shell gravitons in a static spherically symmetric setup of $d$-dimensional spacetime described by the ansatz metric,
\begin{equation} \label{metric}
    ds^2 = -L(r)^2 dt^2 + \frac{dr^2}{\xi(r)} + r^2 d\Omega_{d - 2}^2,
\end{equation}
where $d\Omega_{d - 2}^2$ is the metric of $(d - 2)$-dimensional compact smooth manifold $\M$ and the functions $L(r)$ and $\xi(r)$ are arbitrary. This condensate is composed of $N$ off-shell gravitons with wavelength $\lambda$ localized in the region with radius $R$ and $\widetilde{N}$ off-shell gravitons with wavelength $\widetilde{\lambda}$ in the region with characteristic length scale $\widetilde{R} > R$. The former region will form an object that looks like a black hole as seen by outside observers in the semiclassical limit, while the latter one will become a curved exterior background.

Since the gravitons are weakly coupled, we have $\alpha < {\hbar G_d}/{\lambda^{d - 2}} = {L_p^{d - 2}}/{\lambda^{d - 2}}$, where the latter is the characteristic dimensionless quantum self-coupling of gravitons, $G_d$ is the $d$-dimensional gravitational constant, and $L_p$ is the Planck length. If $M$ is the mass of the condensate, then the number of gravitons is given by $N \sim M/{({\hbar}/{\lambda})} = {G_d M \lambda}/{L_P^{d - 2}}$. The typical value of graviton wavelength is $\lambda \sim (G_d M)^{1/(d - 3)}$, which implies $\alpha N < 1$. It means that we are working in the regime far from the critical point of a quantum phase transition which occurs at $\alpha N = 1$. As a first approximation, we will use the generalized Einstein-Hilbert action for $f(\R)$ gravity as the effective gravitational action for this condensate,
\begin{equation}
    S = \frac{1}{16 \pi G_d} \int d^dx \, \sqrt{-g} f(\R),
\end{equation}
where $g$ is the determinant of the metric in Eq.~\eqref{metric} and $\R$ is the $d$-dimensional Ricci scalar curvature. (For a discussion of condensate of off-shell gravitons described by the four-dimensional action of ordinary gravity featuring nonlocal gravitational interaction, see Ref.~\cite{Buoninfante-Mazumdar-nonlocal-2019}.) Using the ADM formalism \cite{Arnowitt-et-al-dynamics-1962}, we find that the gravitational Hamiltonian of the condensate takes the form \cite{Gao-modified-2010}
\begin{equation} \label{gravitational-Hamiltonian}
    H = -\frac{\Omega_{d - 2}}{16 \pi G_d} \int dr \, r^{d - 2} \frac{L(r)}{\sqrt{\xi(r)}} f{\left( {}^{(d - 1)}\R \right)},
\end{equation}
with $\Omega_{d - 2}$ the volume of $\M$. If $\M$ is a $(d - 2)$-dimensional sphere $S^{d-2}$, then $\Omega_{d - 2} = {2 \pi^{(d - 1)/2}}/{\Gamma{\left( \frac{d - 1}{2} \right)}}$, where $\Gamma$ is the gamma function. The $(d - 1)$-dimensional Ricci scalar curvature ${}^{(d - 1)}\R$ has the form
\begin{equation} \label{Ricci-expression}
    {}^{(d - 1)}\R = -\frac{(d - 2)}{r^2} \left[ \xi'(r) r + (d - 3) \xi(r) \right] + \frac{{}^{(d - 2)}\R}{r^2},
\end{equation}
where ${}^{(d - 2)}\R$ is the Ricci scalar curvature of $\M$, which is taken from now on to be constant, ${}^{(d - 2)}\R \equiv A_0$. In the ordinary case where $\M = S^{d - 2}$, then $A_0 = (d - 2) (d - 3)$.

Following Ref.~\cite{Cunillera-Germani-gross-2018}, we want to study this condensate using the GP equations, which are usually derived in the case of ordinary BEC by making use of a variational approach, namely by minimizing the energy of the condensate while the particle number is kept fixed (see, for example, Refs.~\cite{Pitaevskii-Stringari-bose-2003,Pethick-Smith-bose-2008}). We start with the gravitational energy of our condensate as given in Eq.~\eqref{gravitational-Hamiltonian} and minimize this energy functional with a constraint that the number of gravitons is constant. Therefore, we need to vary the functional $H - \mu N$, where $\mu$ is the chemical potential which serves as a Lagrange multiplier. It is important to emphasize that we have used an off-shell formulation here; the on-shell Einstein equations are not assumed to be satisfied.

The number of off-shell gravitons $N$ can be found from the relation $N = \langle E \rangle \lambda$ where $\langle E \rangle$ is the spatial average of the energy of the gravitons \cite{Cunillera-Germani-gross-2018}. Due to the gravitational redshift, the energy $E$ is related to the energy measured at infinity $E_\infty$ through the relation
\begin{equation} \label{energy-tolman}
    E(r) = L(r) E_\infty.
\end{equation}
As in Ref.~\cite{Cunillera-Germani-gross-2018}, the chemical potential can be written as $\mu = -\gamma/\lambda$ for a constant $\gamma$. Therefore, the term $\mu N$ takes the form
\begin{equation}
    \mu N = -\beta^2 \int dr \, r^{d - 2} \frac{L(r)}{\sqrt{\xi(r)}},
\end{equation}
where all constants are absorbed to $\beta^2$.

\begin{figure}[t]
    \centering
    \includegraphics[width=\columnwidth]{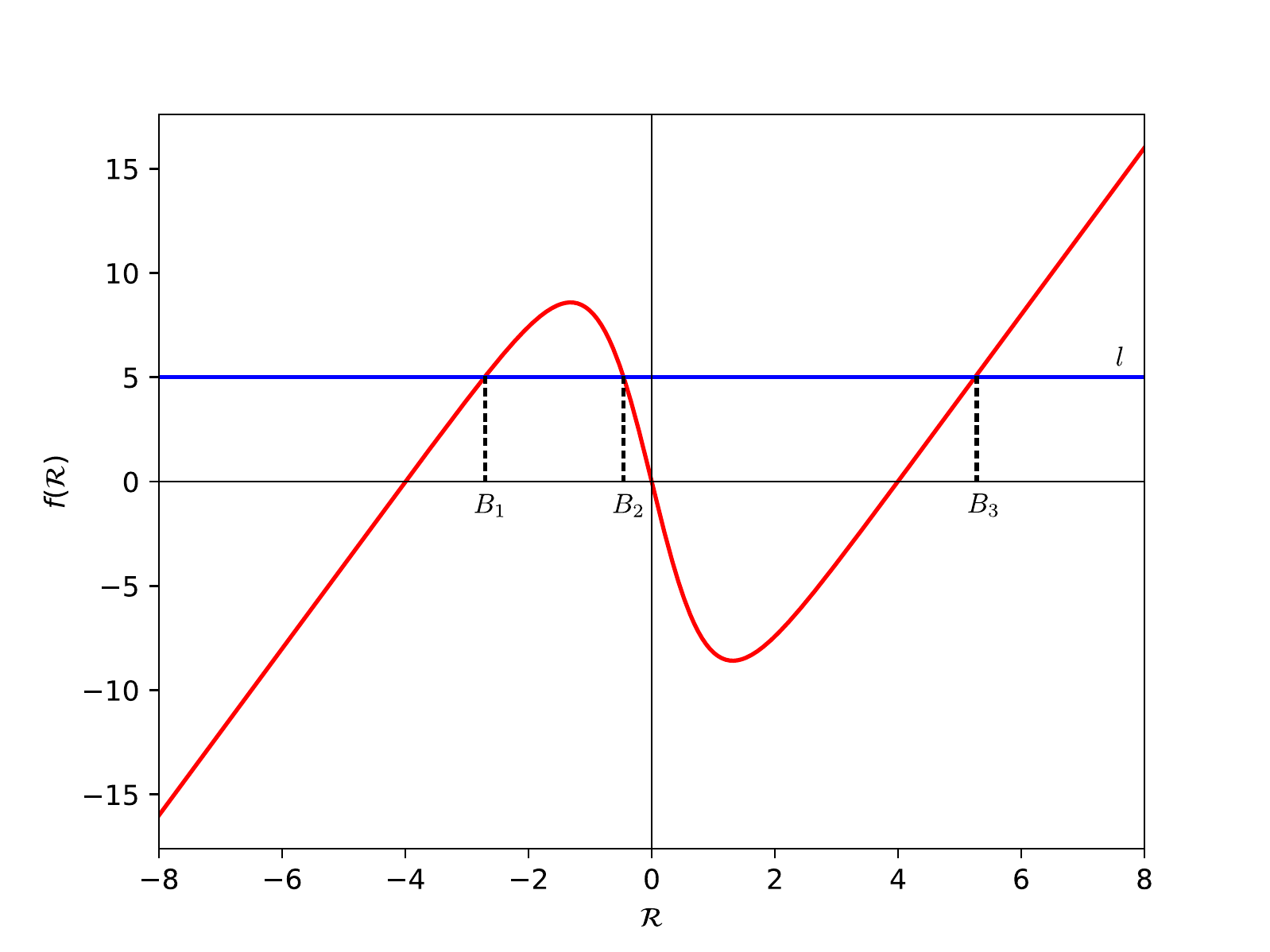}
    \caption{An illustration of finding the constant $B$ for the hyperbolic model of the $f(\R)$ gravity, $f(\R) = \R - b \R_0 \tanh{({\R}/{\R_0})}$, with constants $b, \R_0 > 0$ \cite{Tsujikawa-observational-2008}. Since $B$ is the root of the equation $f(\R) - l = 0$, we have three possible solutions for $B$ in the figure above: two solutions correspond to AdS ($B_1, B_2 < 0$) and one solution corresponds to dS ($B_3 > 0$). However, note that $f'(B_2) < 0$, hence the solution $B_2$ should be ruled out. If we work in ordinary gravity where $f(\R) = \R$ and expect $l > 0$ for the interior and $l = 0$ for the exterior, the only possible solution is dS for the interior spacetime ($B > 0$) and Schwarzschild for the exterior spacetime ($B = 0$), as in Ref.~\cite{Cunillera-Germani-gross-2018}.}
    \label{fig-hypmodel}
\end{figure}

Performing the variation of $H - \mu N$ with respect to $L(r)$ gives us the first GP equation,
\begin{equation} \label{first-GP}
    f{\left( {}^{(d - 1)}\R \right)} = l,
\end{equation}
with $l \equiv 16 \pi G_d \beta^2/{\Omega_{d - 2}}$. It means that ${}^{(d - 1)}\R$ is constant,
\begin{equation} \label{Ricci-B}
    {}^{(d - 1)}\R = B,
\end{equation}
where $B$ is the root of the equation $f\left( {}^{(d - 1)}\R \right) - l = 0$ (see Fig.~\ref{fig-hypmodel}). Therefore, we are always dealing with a space of constant $(d - 1)$-dimensional Ricci scalar curvature ${}^{(d - 1)}\R$ no matter which model of $f(\R)$ gravity that we choose. By substituting Eq.~\eqref{Ricci-expression} to Eq.~\eqref{Ricci-B}, we find that the function $\xi(r)$ takes the form
\begin{equation}
    \xi(r) = \frac{A_0}{(d - 2) (d - 3)} - \frac{B}{(d - 1) (d - 2)} r^2 - \frac{C}{r^{d - 3}},
\end{equation}
where $C$ is an integration constant. For the interior solution, we need to set $C_i = 0$ to ensure a regular solution at the origin $r = 0$.

Performing the variation of $H - \mu N$ with respect to $\xi(r)$ gives us the second GP equation,
\begin{widetext}
\begin{equation} \label{second-GP}
    -\frac{\Omega_{d - 2}}{16 \pi G_d} \Bigg\{ 2 (d - 2) \xi(r) \frac{d}{dr} \left[ f'{\left( {}^{(d - 1)}\R \right)} L(r) \right] - (d - 2) f'{\left( {}^{(d - 1)}\R \right)} L(r) \xi'(r) - f{\left( {}^{(d - 1)}\R \right)} L(r) r \Bigg\} - \beta^2 L(r) r = 0,
\end{equation}
\end{widetext}
where the primed function always denotes its derivative with respect to its argument. Substituting Eq.~\eqref{first-GP} to Eq.~\eqref{second-GP} and solving the resulting equation, we find
\begin{equation}
    L(r) = \sqrt{\xi(r)},
\end{equation}
up to a proportionality constant which can be absorbed to the time parameter in the metric. Therefore, the interior solution can be dS ($B_i > 0$) or AdS ($B_i < 0$), while the exterior can be Schwarzschild ($B_e = 0$), Sch-dS ($B_e > 0$), or Sch-AdS ($B_e < 0$), provided $C_e > 0$, which will be assumed throughout the remainder of this paper. This is in contrast to the system discussed in Ref.~\cite{Cunillera-Germani-gross-2018}, where the gravitons are localized only inside the region $R$ and the ordinary gravity $f(\R) = \R$ is used. Expecting $l > 0$ for the interior and $l = 0$ for the exterior, the solution is therefore always dS for the interior spacetime ($B_i > 0$) and always Schwarzschild for the exterior spacetime ($B_e = 0$).

\section{The size of the interior region}

\label{size-interior-condensate}

The size $R$ of the interior region can be determined using two considerations. First is by matching the interior and exterior solutions at $r = R$ to ensure the continuity at the boundary, namely $\xi_i(R) = \xi_e(R)$. It yields
\begin{equation} \label{continuity-condition}
    C_e = \frac{B_i - B_e}{(d - 1) (d - 2)} R^{d - 1}.
\end{equation}
From the equation above it is clear that the condition $C_e > 0$ puts a restriction $B_i > B_e$. If we interpret $B/2$ as the cosmological constant, it means that the cosmological constant for the exterior must be smaller than the one for the interior, as in Ref.~\cite{Chan-et-al-how-2009}. Hence, we obtain:
\begin{enumerate}[(i)]
    \item If the interior is dS ($B_i > 0$), then the exterior can be Sch-dS ($B_e > 0$), Schwarzschild ($B_e = 0$), or Sch-AdS ($B_e < 0$).
    \item If the interior is AdS ($B_i < 0$), then the exterior must be Sch-AdS ($B_e < 0$).
\end{enumerate}

The second consideration to determine the size $R$ is by matching the energy due to the Gibbons-Hawking-York boundary term $E_{\text{GHY}}$ \cite{Hawking-Horowitz-gravitational-1996} with the energy in Eq.~\eqref{energy-tolman} evaluated at $r = R$, namely $E(R) = E_\infty \sqrt{\xi_i(R)}$. We identify the energy $E_\infty$ as the Komar mass, which is the same for the case of Schwarzschild, Sch-dS, and Sch-AdS \cite{Gunara-et-al-higher-2018}. Therefore, $E_\infty = {(d - 2) \Omega_{d - 2} C_e}/{(16 \pi G_d)}$. The energy $E_{\text{GHY}}$ in $f(\R)$ gravity takes the form \cite{Guarnizo-et-al-boundary-2010}
\begin{equation}
    E_{\text{GHY}} = \frac{1}{8 \pi G_d} \oint_{\M} d^{d - 2}x \sqrt{h} f'{\left( {}^{(d - 1)}\R \right)} K,
\end{equation}
where $h$ is the determinant of the induced metric on $\M$ and $K$ is the trace of the extrinsic curvature on $\M$. For our case, $E_{\text{GHY}}$ is given by
\begin{equation}
    E_{\text{GHY}} = \frac{(d - 2) \Omega_{d - 2}}{8 \pi G_d} f'{\left( B_i \right)} R \sqrt{\xi_i(R)},
\end{equation}
where $f'{\left( B_i \right)}$ is the first derivative $f'{\left( {}^{(d - 1)}\R \right)}$ evaluated at ${}^{(d - 1)}\R = B_i$. Requiring $E_{\text{GHY}} = E(R)$, we find
\begin{equation}
    \left[ \frac{(d - 2) \Omega_{d - 2}}{8 \pi G_d} f'{\left( B_i \right)} R - E_\infty \right] \sqrt{\xi_i(R)} = 0.
\end{equation}
This equation gives us two possible values for $R$,
\begin{eqnarray}
    R &=& s_i \label{first-R1}, \\
    R &=& \frac{C_e}{2 f'{\left( B_i \right)}}, \label{second-R1}
\end{eqnarray}
where $s_i$ is the horizon of the interior solution, $\xi_i(s_i) = 0$. Since we assume $C_e > 0$ and expect $R > 0$, we find that $f'(B_i) > 0$. Hence, the solution $B_i$ with $f'(B_i) < 0$ should be ruled out, such as the root $B_2$ in Fig.~\ref{fig-hypmodel}.

\section{Special cases for the interior and exterior spacetimes}

\label{special-cases}

\subsection{dS interior and Schwarzschild exterior spacetimes}

Let us first consider the case where the interior is dS ($B_i > 0$) and the exterior is Schwarzschild ($B_e = 0$). The continuity condition at the boundary, Eq.~\eqref{continuity-condition}, now reads
\begin{equation}
    C_e = \frac{B_i}{(d - 1) (d - 2)} R^{d - 1}.
\end{equation}
If $s_i$ and $s_e$ are the horizons of the interior and exterior solutions, respectively, which satisfy $\xi_i(s_i) = 0$ and $\xi_e(s_e) = 0$, then
\begin{eqnarray}
    s_i &=& \sqrt{\frac{(d - 1) A_0}{(d - 3) B_i}}, \label{interior-horizon} \\
    s_e &=& \left[ \frac{(d - 2) (d - 3)}{A_0} C_e \right]^{1/{(d - 3)}}.
\end{eqnarray}
From these three equations, we get
\begin{equation} \label{radius-equation-dS-Sch}
    s_i^2 s_e^{d - 3} = R^{d - 1}.
\end{equation}
Choosing the first possible value for $R$ from the $E_{\text{GHY}} = E(R)$ requirement, namely Eq.~\eqref{first-R1}, we obtain $R = s_i = s_e$, which tells us that there is no horizon formation. Therefore, the effective geometry of this graviton condensate is analogous to the gravastar picture, as has been demonstrated previously in Ref.~\cite{Cunillera-Germani-gross-2018}.

\begin{figure}[t]
    \centering
    \includegraphics[width=\columnwidth]{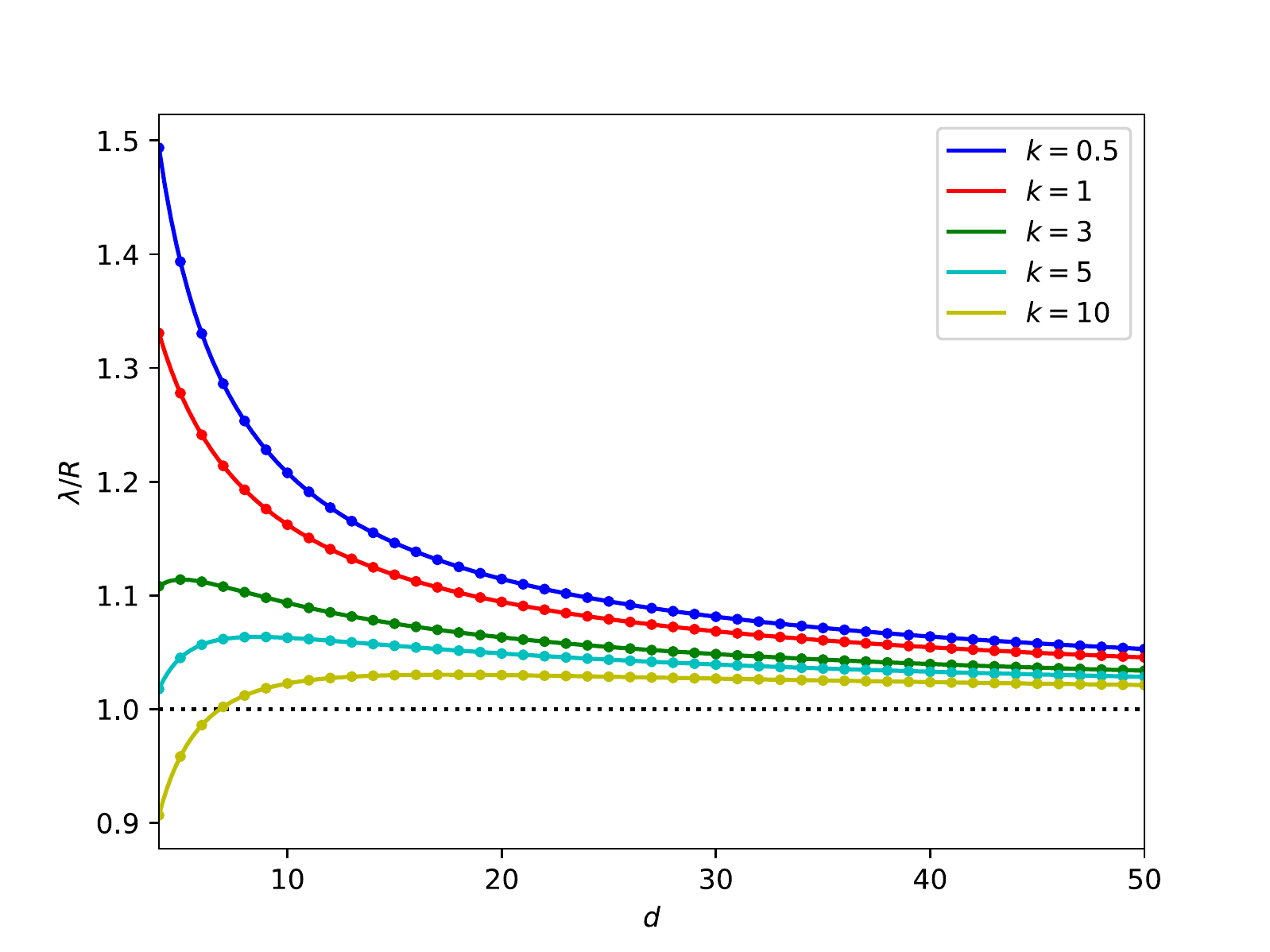}
    \caption{Plot of the ratio between the graviton wavelength $\lambda$ and the size $R$ of the interior versus the spacetime dimension $d$ in the case of general $f(\R)$ gravity for dS interior with Schwarzschild, Sch-dS, and Sch-AdS exteriors when the value $R = s_i = s_e$ is chosen. Note that $\lambda \to R$ for $d \gg$. Here the wavelength is always real valued, which indicates that when the interior geometry is dS spacetime it is always possible for the radius $R$ to have the value $R = s_i = s_e$ in the case of general $f(\R)$ gravity.}
    \label{fig-lambdadSfirstR1}
\end{figure}

We also need to require that the volume of the interior region is equal to $\frac{\Omega_{d - 2}}{d - 1} \lambda^{d - 1}$ \cite{Cunillera-Germani-gross-2018}. Mathematically,
\begin{equation}
    \Omega_{d - 2} \int_0^R dr \frac{r^{d - 2}}{\sqrt{\xi_i(r)}} = \frac{\Omega_{d - 2}}{d - 1} \lambda^{d - 1},
\end{equation}
which then yields
\begin{equation} \label{lambda-1-dS}
    \left( \frac{\lambda}{R} \right)^{d - 1} = \sqrt{\frac{(d - 2) (d - 3)}{A_0}} \, {}_2F_1{\left( \frac{1}{2}, \frac{d - 1}{2}; \frac{d + 1}{2}; \frac{R^2}{s_i^2} \right)},
\end{equation}
where ${}_2F_1$ is the hypergeometric function. As discussed above, here we want to set $R = s_i$. Note that $\lambda \to R$ for very large spacetime dimension $d \gg$. Throughout the remainder of this paper, we will assume that $\M$ is a maximally symmetric space, such that $A_0 = k (d - 2) (d - 3)$, with a constant $k > 0$. The case $k = 1$ is where $\M = S^{d - 2}$, in which we get the value $\lambda = 1.33 \, R$ for $d = 4$. We plot the ratio $\lambda/R$ versus the spacetime dimension $d$ with various values of $k$ for the case of general $f(\R)$ gravity in Fig.~\ref{fig-lambdadSfirstR1}.

If we choose the second possible value for $R$ as in Eq.~\eqref{second-R1}, then from the continuity condition at the boundary we find an equation that can be used to determine the value of $C_e$ given the values of $B_i$ and $f'(B_i)$,
\begin{equation}
    1 = \frac{B_i}{(d - 1) (d - 2)} \frac{C_e^{d - 2}}{\left[ 2 f'(B_i) \right]^{d - 1}}.
\end{equation}
This solution can be identical to or distinct from the previous case while still preventing the horizon formation. Therefore, we require that the horizon of the exterior to be equal to or smaller than $R$ so that, using Eq.~\eqref{radius-equation-dS-Sch}, the horizon of the interior is automatically equal to or larger than $R$, namely $s_e \leq R \leq s_i$. It yields
\begin{equation}
    \left[ 2 f'(B_i) \right]^{d - 3} \leq \frac{A_0}{(d - 2) (d - 3)} C_e^{d - 4}.
\end{equation}
Combined with the previous equation, we arrive at the inequality
\begin{equation} \label{bound-dS-Sch}
    f'(B_i) \leq f'(B_i)_{\text{max}},
\end{equation}
where
\begin{equation}
    f'(B_i)_{\text{max}} = \frac{1}{2} \frac{A_0}{(d - 2) (d - 3)} \left[ \frac{(d - 1) A_0}{(d - 3) B_i} \right]^{\frac{d}{2} - 2}.
\end{equation}
If this inequality is strictly satisfied, then it is possible for the radius $R$ to have a value $R = {C_e}/{[2 f'(B_i)]}$ as in Eq.~\eqref{second-R1} that is distinct from the value $s_i$ as in Eq.~\eqref{first-R1} and that the relation $s_e < R < s_i$ holds. If this inequality is saturated, then the two possible values for $R$ in Eq.~\eqref{first-R1} and \eqref{second-R1} are identical, namely $R = s_i = s_e = {C_e}/{[2 f'(B_i)]}$. However, if this inequality is not satisfied, then it is not possible for the radius $R$ to have a value $R = {C_e}/{[2 f'(B_i)]}$ as in Eq.~\eqref{second-R1}, so the only possible value for $R$ is Eq.~\eqref{first-R1}, namely $R = s_i = s_e$.

\begin{figure}[t]
    \centering
    \includegraphics[width=\columnwidth]{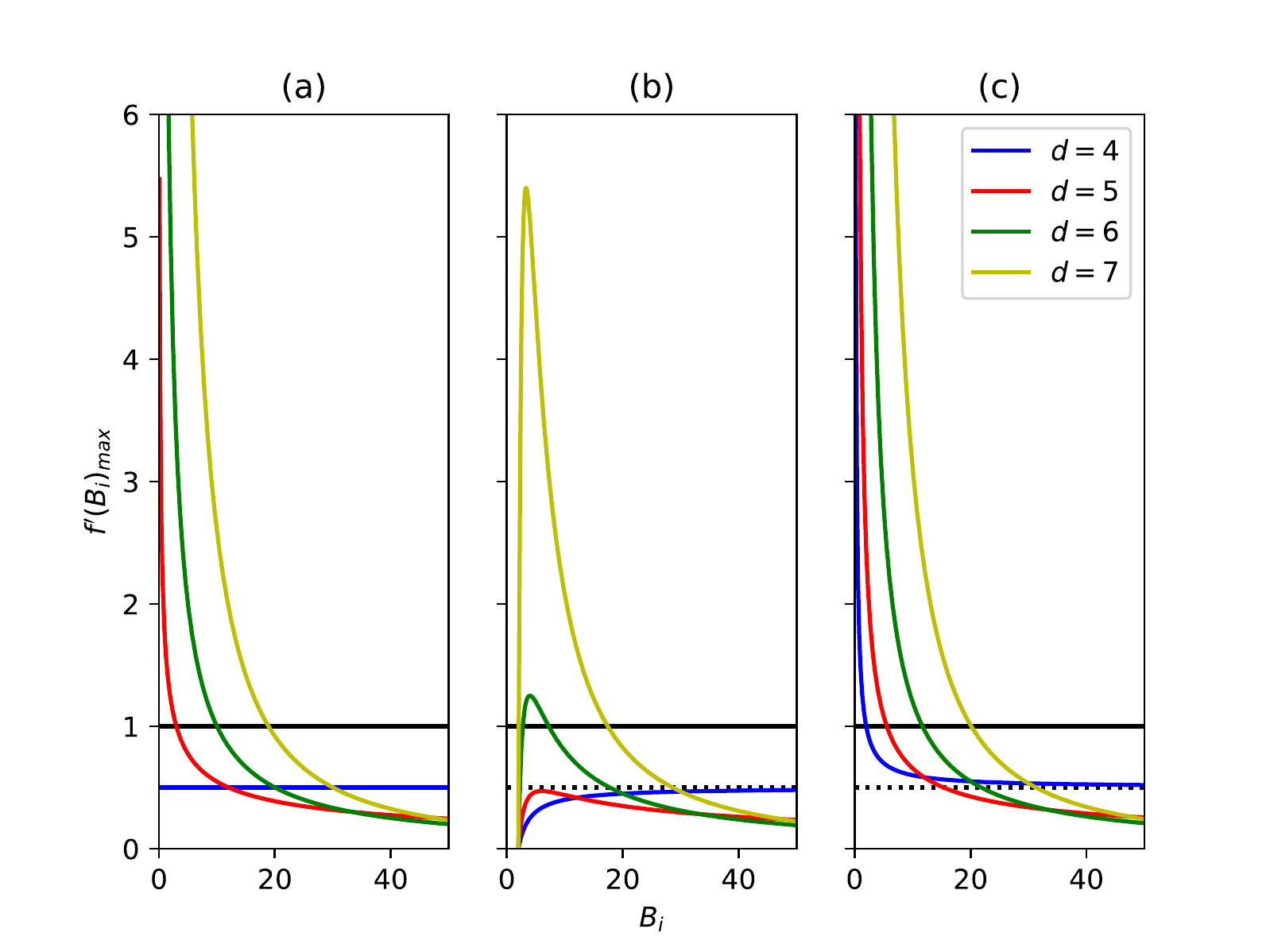}
    \caption{Plot of $f'(B_i)_{\text{max}}$ as a function of $B_i$ for various spacetime dimensions $d$, in the case of dS interior with: (a) Schwarzschild exterior, (b) Sch-dS exterior (where we set $B_e = 2$), and (c) Sch-AdS exterior (with $B_e = -2$), when $R$ is chosen to have a value as in Eq.~\eqref{second-R1}. Here we set $\M = S^{d - 2}$, such that $k = 1$. The ordinary gravity where $f'(B_i) = 1$ for all values of $B_i$ is shown in solid black line. We find that for $d = 4$, $f'(B_i)_{\text{max}}$ is constant with value $k/2$ in the case of Schwarzschild exterior and monotonically increasing (decreasing) to the asymptotic value $k/2$ in the case of Sch-(A)dS exterior. For dimension $d > 4$, $f'(B_i)_{\text{max}}$ is monotonically decreasing to zero as $B_i \to \infty$ in the case of Schwarzschild and Sch-AdS exteriors, but displays a nonmonotonic behavior in the case of Sch-dS exterior and approaches zero as $B_i \to \infty$.}
    \label{fig-f1BmaxdS}
\end{figure}

If we set $d = 4$, we find that $f'(B_i)_{\text{max}}$ is constant with value $k/2$, independent of the value of $B_i$, while if we set $d > 4$ and an arbitrary value of $k$, $f'(B_i)_{\text{max}}$ is monotonically decreasing to zero as $B_i \to \infty$ [see Fig.~\ref{fig-f1BmaxdS}(a)]. If we are working in the ordinary gravity, where $f'(B_i) = 1$ for all values of $B_i$, and considering the case $d = 4$, then for $k > 2$ the inequality in Eq.~\eqref{bound-dS-Sch} is strictly satisfied, which implies that the radius $R$ can have a value $R = {C_e}/{[2 f'(B_i)]}$ that is distinct from $s_i$ and that $s_e < R < s_i$. If $k = 2$, the inequality is saturated, so that $R = s_i = s_e = {C_e}/{[2 f'(B_i)]}$. If $k < 2$, which also includes the special case $k = 1$ where $\M = S^{d - 2}$, then the inequality is not satisfied, so that the only possibility is $R = s_i = s_e$, as in Ref.~\cite{Cunillera-Germani-gross-2018}. For the case $d > 4$ and arbitrary value of $k$, there exists a critical value $\zeta_{\text{Sch}}(d > 4; k)$ which satisfies $f'(\zeta_{\text{Sch}}(d > 4; k)) = 1$, namely
\begin{equation}
    \zeta_{\text{Sch}}(d > 4; k) = \frac{(d - 1) (d - 2)}{2^{2/(d - 4)}} k^{(d - 2)/(d - 4)},
\end{equation}
such that the inequality in Eq.~\eqref{bound-dS-Sch} is strictly satisfied in the regime $0 < B_i < \zeta_{\text{Sch}}(d > 4; k)$, saturated at $B_i = \zeta_{\text{Sch}}(d > 4; k)$, and not satisfied in the regime $B_i > \zeta_{\text{Sch}}(d > 4; k)$. The consequences in terms of the possible values of $R$ for each of these regimes are again the same as above.

The graviton wavelength in the case of general $f(\R)$ gravity, when the value $R = {C_e}/{[2 f'(B_i)]}$ is chosen, is again given by Eq.~\eqref{lambda-1-dS}, with the value of ${R^2}/{s_i^2}$ now becomes
\begin{equation}
    \frac{R^2}{s_i^2} = \left[ \frac{f'(B_i)}{f'(B_i)_{\text{max}}} \right]^{\frac{2}{d - 2}}.
\end{equation}
Therefore, we find
\begin{eqnarray} \label{lambda-1-dS-second-value}
    \left( \frac{\lambda}{R} \right)^{d - 1} &=& \sqrt{\frac{(d - 2) (d - 3)}{A_0}} \times \\
    && {}_2F_1{\left( \frac{1}{2}, \frac{d - 1}{2}; \frac{d + 1}{2}; \left[ \frac{f'(B_i)}{f'(B_i)_{\text{max}}} \right]^{\frac{2}{d - 2}} \right)} \nonumber.
\end{eqnarray}
For very large spacetime dimension $d \gg$, we again have $\lambda \to R$. In the regime where the inequality in Eq.~\eqref{bound-dS-Sch} is not satisfied, the wavelength takes unphysical complex values, which indicates that the solution in Eq.~\eqref{second-R1} is not possible. Otherwise, in the regime where the inequality is satisfied, then the wavelength is real valued such that the solution in Eq.~\eqref{second-R1} is possible, which can be identical to or distinct from the solution in Eq.~\eqref{first-R1} depending on whether the inequality is saturated or strictly satisfied, respectively. We plot the ratio $\lambda/R$ versus the spacetime dimension $d$ with various values of $k$ for the case of ordinary gravity in Fig.~\ref{fig-lambdadSsecondR1} (dots in solid lines).

\subsection{dS interior and Sch-(A)dS exterior spacetimes}

Now we will discuss the case of dS interior where the exterior spacetime can be Sch-dS or Sch-AdS. From the expression for the horizon of the interior solution $s_i$ given in Eq.~\eqref{interior-horizon} and choosing the first solution $R = s_i$ as in Eq.~\eqref{first-R1}, we obtain
\begin{equation}
    R = \sqrt{\frac{(d - 1) A_0}{(d - 3) B_i}}.
\end{equation}
Inserting this expression to the continuity condition at the boundary given in Eq.~\eqref{continuity-condition} yields
\begin{equation}
    \frac{A_0}{(d - 2) (d - 3)} - \frac{B_e}{(d - 1) (d - 2)} R^2 - \frac{C_e}{R^{d - 3}} = 0,
\end{equation}
which is essentially a statement that $\xi_e(R) = 0$. Therefore, we again find that $R = s_i = s_e$, which, in the case of Sch-dS, $s_e$ means the smaller positive horizon. As before, this indicates that there is no horizon formation. The graviton wavelength $\lambda$ is given by Eq.~\eqref{lambda-1-dS} with ${R^2}/{s_i^2} = 1$, hence in this case the plot of $\lambda/R$ as a function of $d$ will be identical to Fig.~\ref{fig-lambdadSfirstR1}. Since the wavelength in this case is always real valued, together with the result of the previous section, we conclude that when the interior geometry is dS spacetime, it is always possible for the radius $R$ to have the value $R = s_i = s_e$ in the case of general $f(\R)$ gravity.

\begin{figure}[t]
    \centering
    \includegraphics[width=\columnwidth]{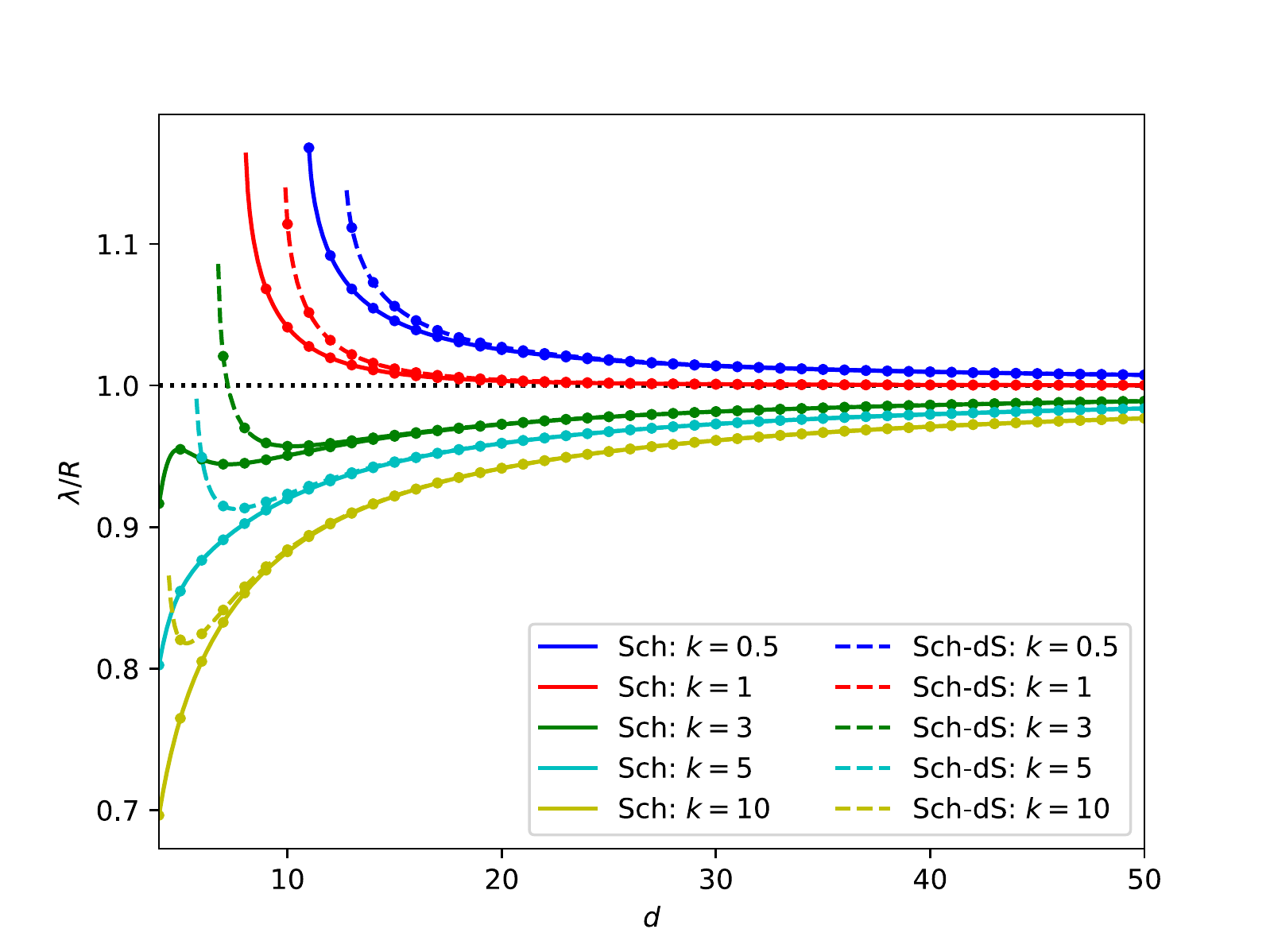}
    \caption{Plot of the ratio between the graviton wavelength $\lambda$ and the size $R$ of the interior versus the spacetime dimension $d$ in the case of ordinary gravity for dS interior with Schwarzschild (dots in solid lines, with $B_i = 30$ and $B_e = 0$) and Sch-dS (dots in dashed lines, with $B_i = 30$ and $B_e = 25$) exteriors when the value $R = {C_e}/{[2 f'(B_i)]}$ is chosen. Note that $\lambda \to R$ for $d \gg$. For the case of general $f(\R)$ gravity, therefore including the ordinary gravity used here in this plot, when the radius $R$ has the value $R = {C_e}/{[2 f'(B_i)]}$ there is an inequality restriction $f'(B_i) \leq f'(B_i)_{\text{max}}$, which exists only when the interior geometry is dS spacetime, that has to be satisfied such that the wavelength is real valued. In the regime where this inequality is not satisfied, the wavelength takes unphysical complex values, which indicates that it is not possible in that regime for the radius $R$ to have the value $R = {C_e}/{[2 f'(B_i)]}$.}
    \label{fig-lambdadSsecondR1}
\end{figure}

If we choose the second possible value for $R$ as in Eq.~\eqref{second-R1}, then we require $s_e \leq R$ so that this solution is identical to or distinct from the previous case while still preventing the horizon formation. Using the equation $\xi_e(s_e) = 0$ and the continuity condition at the boundary, we obtain
\begin{equation} \label{R-se-dS-SchdSAdS}
    \left( \frac{R}{s_e} \right)^{d - 1} = \frac{({z_e^2}/{s_e^2}) \pm 1}{({z_e^2}/{s_i^2}) \pm 1},
\end{equation}
with the minus (plus) sign is for the case of Sch-(A)dS exterior, and we have defined
\begin{equation}
    z_{i, e} \equiv \sqrt{\frac{(d - 1) A_0}{(d - 3) |B_{i, e}|}},
\end{equation}
where $z_i$ is simply $s_i$ here. Note that in the case of Sch-dS, $z_e$ is the cosmological horizon of the exterior spacetime. Since $R$ must be positive, for the case of Sch-dS exterior we have $z_e > s_i$ or $B_i > B_e$, which is automatically satisfied. Using Eq.~\eqref{R-se-dS-SchdSAdS}, the requirement $s_e \leq R$ gives us $s_e \leq s_i$.

There are now two possible choices of inequalities, $s_e \leq R \leq s_i$ and $s_e \leq s_i \leq R$, where only the former which will give us physical results. The inequality $R \leq s_i$ implies
\begin{equation} \label{bound-dS-Sch-dSAdS}
    f'(B_i) \leq f'(B_i)_{\text{max}},
\end{equation}
where
\begin{equation} \label{fBmax-dS-AdS}
    f'(B_i)_{\text{max}} = \frac{1}{2} \frac{A_0}{(d - 2) (d - 3)} \left( 1 - \frac{B_e}{B_i} \right) \left[ \frac{(d - 1) A_0}{(d - 3) B_i} \right]^{\frac{d}{2} - 2}.
\end{equation}
We can then apply a similar analysis as in the case of dS interior and Schwarzschild exterior. If we set $d = 4$ and keep $B_e$ fixed, we find that as $B_i$ is varied, $f'(B_i)_{\text{max}}$ is monotonically increasing (decreasing) to the asymptotic value $k/2$ in the case of Sch-(A)dS exterior, while if we set $d > 4$ and an arbitrary value of $k$, $f'(B_i)_{\text{max}}$ is monotonically decreasing to zero as $B_i \to \infty$ in the case of Sch-AdS exterior, but displays a nonmonotonic behavior in the case of Sch-dS exterior and approaches zero as $B_i \to \infty$ [see Fig.~\ref{fig-f1BmaxdS}(b,c)].

If we are working in the ordinary gravity and focusing on the case of Sch-dS exterior at $d = 4$ with $B_e > 0$ kept fixed, then the inequality in Eq.~\eqref{bound-dS-Sch-dSAdS} is not satisfied if $k \leq 2$, and there exists a critical value $\zeta_{\text{Sch-dS}}(d = 4; k > 2)$ if $k > 2$ which satisfies $f'(\zeta_{\text{Sch-dS}}(d = 4; k > 2)) = 1$, namely
\begin{equation} \label{zeta-Sch-dS}
    \zeta_{\text{Sch-dS}}(d = 4; k > 2) = \frac{B_e}{1 - \frac{2}{k}},
\end{equation}
such that the inequality is not satisfied in the regime $B_e < B_i < \zeta_{\text{Sch-dS}}(d = 4; k > 2)$, saturated at $B_i = \zeta_{\text{Sch-dS}}(d = 4; k > 2)$, and strictly satisfied in the regime $B_i > \zeta_{\text{Sch-dS}}(d = 4; k > 2)$. For the case $d > 4$ and arbitrary value of $k$, there may exist two critical values $\zeta_{\text{Sch-dS}}^<(d > 4; k)$ and $\zeta_{\text{Sch-dS}}^>(d > 4; k)$, where the latter is the larger of the two, such that the inequality is not satisfied in the regimes $B_e < B_i < \zeta_{\text{Sch-dS}}^<(d > 4; k)$ and $B_i > \zeta_{\text{Sch-dS}}^>(d > 4; k)$, saturated at $B_i = \zeta_{\text{Sch-dS}}^<(d > 4; k)$ and $B_i = \zeta_{\text{Sch-dS}}^>(d > 4; k)$, and strictly satisfied in the regime $\zeta_{\text{Sch-dS}}^<(d > 4; k) < B_i < \zeta_{\text{Sch-dS}}^>(d > 4; k)$. If these two critical values coincide, then the inequality is not satisfied for all values of $B_i$ except at $B_i = \zeta_{\text{Sch-dS}}^<(d > 4; k) = \zeta_{\text{Sch-dS}}^>(d > 4; k)$ where it is saturated. Otherwise, if there is no critical value, then the inequality is not satisfied for all values of $B_i$ without exception.

Now let us focus on the case of Sch-AdS exterior while still working in the ordinary gravity with $B_e < 0$ kept fixed. At $d = 4$ with $k \geq 2$, the inequality in Eq.~\eqref{bound-dS-Sch-dSAdS} is strictly satisfied. At $d = 4$ with $k < 2$, and also for $d > 4$ with arbitrary value of $k$, there exists a critical value $\zeta_{\text{Sch-AdS}}(d; k)$ which satisfies $f'(\zeta_{\text{Sch-AdS}}(d; k)) = 1$, such that the inequality is strictly satisfied in the regime $0 < B_i < \zeta_{\text{Sch-AdS}}(d; k)$, saturated at $B_i = \zeta_{\text{Sch-AdS}}(d; k)$, and not satisfied in the regime $B_i > \zeta_{\text{Sch-AdS}}(d; k)$. The critical value at $d = 4$ with $k < 2$, namely $\zeta_{\text{Sch-AdS}}(d = 4; k < 2)$, has the same expression as in Eq.~\eqref{zeta-Sch-dS}, but it is important to emphasize that here $B_e < 0$ and $k < 2$.

\begin{figure}[t]
    \centering
    \includegraphics[width=\columnwidth]{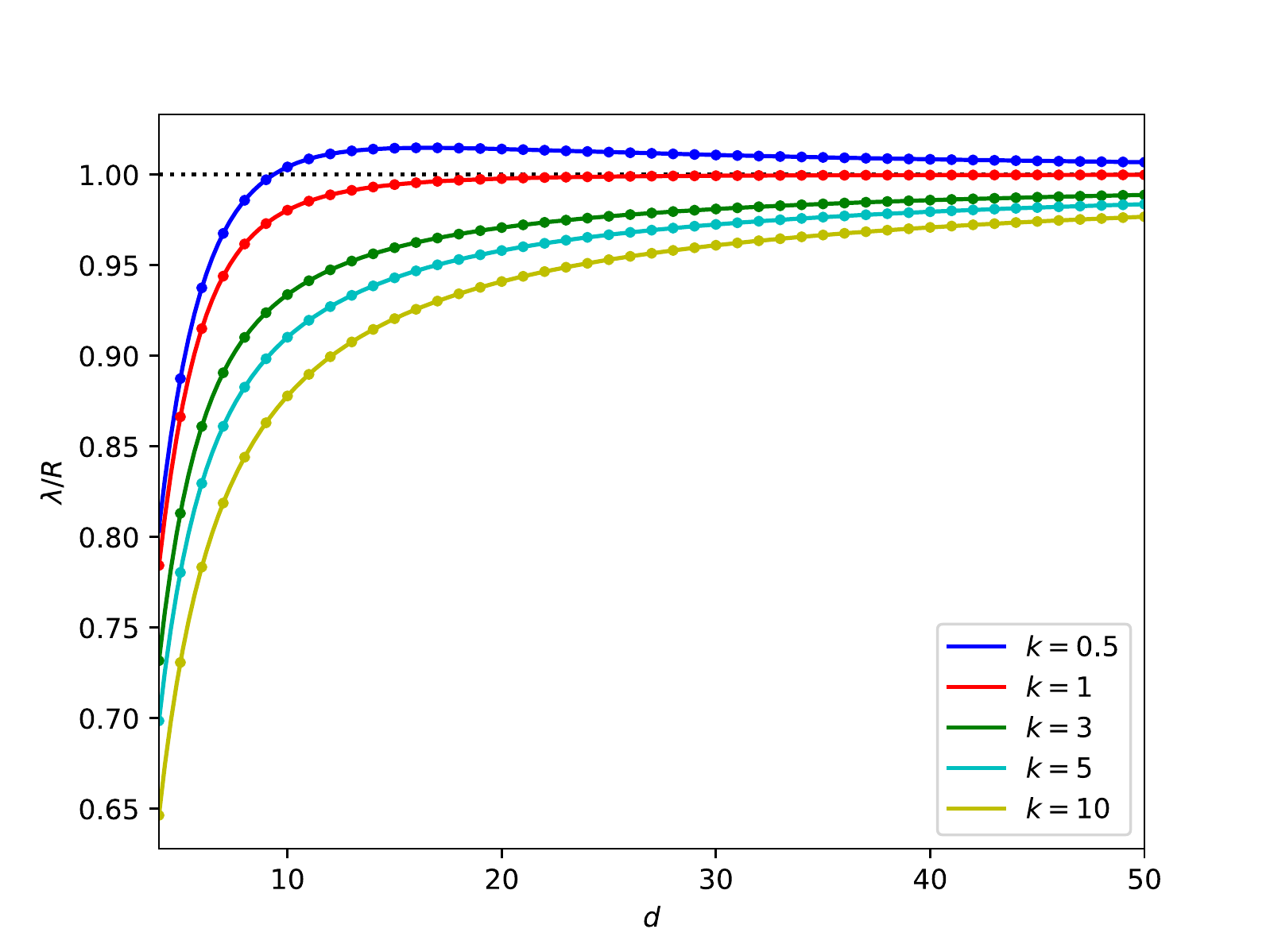}
    \caption{Plot of the ratio between the graviton wavelength $\lambda$ and the size $R$ of the interior versus the spacetime dimension $d$ in the case of ordinary gravity for AdS interior with Sch-AdS exterior (with $B_i = -25$ and $B_e = -30$), where the radius $R$ always has the value $R = {C_e}/{[2 f'(B_i)]}$. Note that $\lambda \to R$ for $d \gg$. For the case of general $f(\R)$ gravity, therefore including the ordinary gravity used here in this plot, there is no inequality restriction in this case that has to be satisfied, in contrast to the case where the interior is dS. This implies that the wavelength is always real valued. Therefore, we conclude that when the interior geometry is AdS spacetime it is always and only possible for the radius $R$ to have the value $R = {C_e}/{[2 f'(B_i)]}$ in the case of general $f(\R)$ gravity.}
    \label{fig-lambdaAdS}
\end{figure}

The graviton wavelength in the case of general $f(\R)$ gravity, when the value $R = {C_e}/{[2 f'(B_i)]}$ is chosen, is given by Eq.~\eqref{lambda-1-dS-second-value}, but now Eq.~\eqref{fBmax-dS-AdS} is used for $f'(B_i)_{\text{max}}$. We again note that $\lambda \to R$ for very large spacetime dimension $d \gg$. Together with the result of the previous section, we conclude that for the case of general $f(\R)$ gravity, when the radius $R$ has the value $R = {C_e}/{[2 f'(B_i)]}$ and the interior geometry is dS spacetime, there is an inequality restriction $f'(B_i) \leq f'(B_i)_{\text{max}}$ that has to be satisfied such that the wavelength is real valued. In the regime where this inequality is not satisfied, the wavelength takes unphysical complex values, which indicates that it is not possible in that regime for the radius $R$ to have the value $R = {C_e}/{[2 f'(B_i)]}$. We plot the ratio $\lambda/R$ versus the spacetime dimension $d$ with various values of $k$ for the case of ordinary gravity in Fig.~\ref{fig-lambdadSsecondR1} (dots in dashed lines).

\subsection{AdS interior and Sch-AdS exterior spacetimes}

If the interior geometry is AdS spacetime, then the exterior must be Sch-AdS due to the inequality $B_i > B_e$. The interior now does not have horizon, so we can only have the value as in Eq.~\eqref{second-R1} for the radius $R$, in contrast to the previous case where the interior is dS. Using the continuity condition at the boundary, Eq.~\eqref{continuity-condition}, and the equation defining $s_e$, namely $\xi_e(s_e) = 0$, the relation between $R$ and $s_e$ can be obtained as
\begin{equation}
    \left( \frac{R}{s_e} \right)^{d - 1} = \frac{1 + ({z_e^2}/{s_e^2})}{1 - ({z_e^2}/{z_i^2})}.
\end{equation}
Since $B_i > B_e$, or $|B_e| > |B_i|$, we find $z_i > z_e$. Therefore, $s_e < R$, which means that in this case there is also no horizon formation. The value of $R$ can be obtained using
\begin{equation}
    R^{d - 2} = \frac{(d - 1) (d - 2)}{|B_e| - |B_i|} \left[ 2 f'(B_i) \right].
\end{equation}
This expression can then be used to calculate the graviton wavelength in the case of general $f(\R)$ gravity, which now takes the form
\begin{equation} \label{lambda-1-AdS}
    \left( \frac{\lambda}{R} \right)^{d - 1} = \sqrt{\frac{(d - 2) (d - 3)}{A_0}} \, {}_2F_1{\left( \frac{1}{2}, \frac{d - 1}{2}; \frac{d + 1}{2}; -\frac{R^2}{z_i^2} \right)}.
\end{equation}
We again note that $\lambda \to R$ for very large spacetime dimension $d \gg$. In contrast to the case where the interior is dS, here there is no inequality restriction that has to be satisfied, which implies that the wavelength is always real valued. Therefore, we conclude that when the interior geometry is AdS spacetime it is always and only possible for the radius $R$ to have the value $R = {C_e}/{[2 f'(B_i)]}$ in the case of general $f(\R)$ gravity. We plot the ratio $\lambda/R$ versus the spacetime dimension $d$ with various values of $k$ for the case of ordinary gravity in Fig.~\ref{fig-lambdaAdS}.

\section{Conclusions}

\label{conclusions}

In this paper, we have studied a model of weakly coupled off-shell gravitons which form Bose-Einstein condensate in the regime that is far from the quantum critical point. We adopted the approach outlined in Ref.~\cite{Cunillera-Germani-gross-2018} while making the following three generalizations: (i) introducing a composite of two sets of off-shell gravitons with different wavelength to enable richer geometries for the interior and exterior spacetimes; (ii) working in $f(\R)$ gravity; and (iii) extending the calculations to higher dimensions. Remaining in the static spherically symmetric setup, we found that the effective spacetime geometry is again analogous to a gravastar, but now with a metric which strongly depends on the function $f(\R)$. This gives more possibilities to the effective spacetime geometries both in the interior and the exterior. The interior spacetime now can be dS or AdS, while the exterior can be Schwarzschild, Sch-dS, or Sch-AdS. These geometries are determined by the function $f(\R)$, unlike in previous works where they were selected by hand. A continuity condition of the metric at the boundary between the interior and exterior regions provides a relation between the interior and exterior spacetimes: the cosmological constant for the exterior must be smaller than the one for the interior.

There are two possible values for the radius $R$ of the interior region, $R = s_i = s_e$ and $R = {C_e}/{[2 f'(B_i)]}$, where the latter is only possible because of the generalizations above. Focusing first on the case where the interior geometry is dS spacetime, such that the exterior geometry can be Schwarzschild, Sch-dS, or Sch-AdS, we found that it is always possible for the radius $R$ to have the value $R = s_i = s_e$. However, when the radius $R$ has the value $R = {C_e}/{[2 f'(B_i)]}$, there is an inequality restriction $f'(B_i) \leq f'(B_i)_{\text{max}}$ that has to be satisfied such that the wavelength is real valued. In the regime where this inequality is not satisfied, the wavelength takes unphysical complex values, which indicates that it is not possible in that regime for the radius $R$ to have the value $R = {C_e}/{[2 f'(B_i)]}$. For the case where the interior geometry is AdS spacetime, such that the exterior geometry is Sch-AdS, we found that the radius $R$ can only have the value $R = {C_e}/{[2 f'(B_i)]}$ and there is no inequality restriction that has to be satisfied, in contrast to the previous case. This implies that the wavelength is always real valued. Therefore, in this case it is always and only possible for the radius $R$ to have the value $R = {C_e}/{[2 f'(B_i)]}$.

\section*{Acknowledgments}

The work in this paper is supported by Riset ITB 2019-2020. B.~E.~G acknowledges the Abdus Salam ICTP for Associateships and for the warmest hospitality where the final part of this paper has been done.

\bibliographystyle{apsrev4-1}
\bibliography{references}

\end{document}